\documentstyle[11pt,fleqn]{article}
\newcommand{\preprint}[1]{\hfill{\sl preprint - #1}\par\bigskip\par\rm}
\def\titolo{\par\bigskip\begin{center}\bf\LARGE}
\def\endtitolo{\end{center}\par\bigskip\par\rm\normalsize}
\def\instit{\begin{center}\large}
\def\endinstit{\end{center}\rm\normalsize}
\def\references{\begin{thebibliography}{10}}
\def\endreferences{\end{thebibliography}\end{document}}
\newcommand{\dip}{\smallskip Dipartimento di Fisica,
                                Universit\`a di Trento, Italia}
\newcommand{\infn}{\smallskip Istituto Nazionale di Fisica Nucleare,\\
                                 Gruppo Collegato di Trento, Italia}
\newcommand{\dinfn}{\dip\\ and \infn}
\newcommand{\btit}{\begin{titolo}}
\newcommand{\etit}{\end{titolo}}

\newcommand{\Idinfn}{\begin{instit}\dinfn\end{instit}}
\newcommand{\s}[1]{\section{#1}\renewcommand{\theequation}
        {\mbox{\arabic{section}.\arabic{equation}}}\setcounter{equation}{0}}

\renewcommand{\author}[1]{\begin{center}\Large #1\end{center}}
\renewcommand{\date}[1]{\par\bigskip\par\sl\hfill #1\par\medskip\par\rm}
\newcommand{\email}[1]{e-mail: \sl #1@science.unitn.it\rm}
\newcommand{\femail}[1]{\footnote{\email{#1}}}
\newcommand{\pacs}[1]{\smallskip\noindent{\sl PACS number(s):
                       \hspace{0.3cm}#1}\par\bigskip\rm}
\newcommand{\babs}{\hrule\par\begin{description}\item{Abstract: }\it}
\newcommand{\eabs}{\par\end{description}\hrule\par\medskip\rm}

\renewcommand{\vec}[1]{{\bf #1}}       
\newcommand{\M}{{\cal M}}              
\newcommand{\hs}{\qquad\qquad}         
\newcommand{\nn}{\nonumber}            
\newcommand{\beq}{\begin{eqnarray}}    
\newcommand{\eeq}{\end{eqnarray}}      
\newcommand{\beqn}{\begin{eqnarray}}   
\newcommand{\eeqn}{\end{eqnarray}}     
\newcommand{\at}{\left(}               
\newcommand{\aq}{\left[}               
\newcommand{\ct}{\right)}              
\newcommand{\cq}{\right]}              
\newcommand{\ii}{\infty}                         
\newcommand{\fr}[2]{\mbox{$\frac{#1}{#2}$}}      
\newcommand{\Tr}{\,\mbox{Tr}\,}                  
\newcommand{\PP}{\,\mbox{PP}\,}                  
\newcommand{\Res}{\,\mbox{Res}\,}                
\renewcommand{\Re}{\,\mbox{Re}\,}                
\newcommand{\lap}{\Delta}                        

\newcommand{\be}{\beta}

\newcommand{\de}{\delta}
\newcommand{\ep}{\varepsilon}
\newcommand{\ze}{\zeta}

\newcommand{\la}{\lambda}

\newcommand{\si}{\sigma}

\newcommand{\ph}{\varphi}

\newcommand{\te}{\vartheta}

\newcommand{\Ga}{\Gamma}

\newcommand{\Om}{\Omega}

\begin{document}

\preprint{UTF 348}
\btit
One-loop Quantum Corrections to the Entropy\\
for an Extremal Reissner-Nordstr\"om Black Hole
\etit
\author{Guido Cognola\femail{cognola}, Luciano Vanzo\femail{vanzo}
and Sergio Zerbini\femail{zerbini}}
\Idinfn

\date{April - 1995}

\babs
The first quantum corrections to the entropy for an eternal 4-dimensional
extremal Reissner-Nordstr\"om black hole is investigated at one-loop level,
in the large mass limit of the black hole,
making use of the conformal  techniques  related to the optical metric.
A leading cubic horizon divergence is found and
other divergences appear due to the singular nature of the optical manifold.
The area law is shown to be violated.
\eabs

\s{Introduction}
Recently, several issues like the interpretation of the
Bekenstein-Hawking classical formula for the black hole entropy, the
loss of information paradox and the validity of the area law have been
discussed in the literature (see, for example, the review
\cite{beke94u-15}). Furthermore it has been pointed out that it should
be desiderable to have the usual statistical interpretation of the
black hole entropy as the number of gravitational states at the
horizon (see for example, \cite{carl94u-32}).

There have also been some attempts to compute
semiclassically the first quantum corrections (prefactor) to the
Bekenstein-Hawking classical entropy  $4\pi G M^2$
\cite{beke73-7-2333,hawk75-43-199}.
With regard to this, we recall that on general grounds, the density of
levels as a function of the black hole mass $M$ for $D=4$ should read
\beq
\Om(M)\simeq C(M)e^{4\pi GM}
\:.\eeq
However, so far all, the evaluations of the prefactor $C(M)$ have been
plagued by the
appearance of  divergences
\cite{thoo85-256-727,bomb86-34-373,suss94-50-2700,dowk94-11-55,solo95-51-609,furs94u-32,empa94u-5}
present also in the related "entanglement or geometric entropy"
\cite{sred93-71-666,call94-333-55,kaba94-329-46}. These divergences
have been related to the information loss issue
\cite{thoo85-256-727,suss94-50-2700} and their presence has to be
confronted with the fact that the corresponding
prefactor for quantum fields and extended objects (like strings or
p-branes) in ultrastatic backgrounds is computable and finite.

In a space-time with horizons, the physical origin of these divergences can
be traced back to the equivalence principle
\cite{isra76-57-107,scia81-30-327,barb94-50-2712}. The argument goes
as follows. A system in thermal equilibrium has a local Tolman
temperature $T(x)=T/\sqrt -g_{00}$, $T$ being the asymptotic
temperature. As a consequence, one has for the entropy $S$ of a quantum
massless field the formal quantity
\beq
S \equiv T^3\int \frac{dV_3}{(-g_{00})^{3/2}}
\eeq
which is divergent if the space-time has horizons and, in turn, suggests that
the use of the optical metric
$\bar{g}_{\mu\nu}=g_{\mu\nu}/|g_{00}|$, conformally
related to the original one,
may provide an alternative and useful framework to
investigate these issues.  One of the purposes of this paper is to
implement this idea, which has been already proposed in Refs.
\cite{page82-25-1499,brow85-31-2514} and put recently forward also in
Refs.~\cite{frol94u-36,barv94u-36,empa94u-5,deal95u-33}.

In Ref. \cite{cogn95u-342}, in the large black hole mass limit, we have
reconsidered the case of an ordinary black
hole admitting a canonical horizon, namely the case for which the
quantity $g_{00}$ has a simple zero.
In this paper, we would like to investigate the
case of an extremal Reissner-Nordstr\"om black hole, within the same
approximation, which is reliable very near the horizon. Here the
quantity $g_{00}$ has a double zero and  this corresponds to
a non canonical horizon. It is known that in this case
the Hawking temperature is vanishing and thus a
vanishing Hawking-Bekenstein entropy is present,
but the horizon area is not vanishing at all \cite{hawk94u-13}.
For this reason, it may be interesting to study the first quantum
corrections to the entropy.

The content of the paper is as follows.
In Sec. \ref{S:conf} we resume the conformal transformation techniques
and give the relation between the the physical quantities in the
original manifolds and in the optical one.
In Sec.~\ref{S:reiss} we derive, in the large mass limit, eigenfunctions and
eigenvalues for a
scalar field in an extremal Reissner-Nordstr\"om background and,
using $\zeta$-function regularization, we derive the free energy and
the entropy, regularizing the horizon divergences with a suitable cutoff.
Finally in Sec.~\ref{S:conc} we conclude with some remarks.

\s{Conformal transformation techniques}
\label{S:conf}

To start with, we recall the formalism we shall use in the following in
order to discuss the finite temperature effects. We may consider, as a
prototype of the quantum correction, a scalar field
on a 4-dimensional static space-time defined by the metric (signature
$-+++$)
\beq
ds^2=g_{00}(\vec{x})(dx^0)^2+g_{ij}(\vec{x})dx^idx^j\:,
\hs \vec{x}=\{x^j\}\:,\hs i,j=1,...,3\:.
\eeq

The one-loop partition function is given by (we perform the Wick
rotation $x_0=-i\tau$, thus all differential operators one is
dealing with will be elliptic)
\begin{equation}
Z=\int d[\phi]\,
\exp\at-\frac12\int\phi L_4 \phi d^4x\ct
\:,\end{equation}
where $\phi$ is a scalar density of weight $-1/2$ and
$L_4$ is a Laplace-like operator on a 4-dimensional manifold.
It has the form
\beq
L_4=-\lap_4+m^2+\xi R
\:.\eeq
Here $\lap_4$ is the Laplace-Beltrami operator, $m$ (the mass)
and $\xi$ arbitrary parameters and $R$ the scalar curvature of the manifold.

The key idea is to make use of the conformal transformation technique.
\cite{dowk78-11-895,gusy87-46-1097,dowk88-38-3327,dowk89-327-267}.
This method is useful because it permits to compute all physical
quantities in an ultrastatic manifold
(called the optical manifold \cite{gibb78-358-467}) and,
at the end of calculations, to transform  back them to a static one,
with an arbitrary $g_{00}$.
The ultrastatic Euclidean metric $\bar{g}_{\mu\nu}$ is related to the static
one by the conformal transformation
\beq
\bar{g}_{\mu\nu}(\vec{x})=e^{2\si(\vec{x})}g_{\mu\nu}(\vec{x})
\:,\eeq
with
$\si(\vec{x})=-\frac{1}{2}\ln g_{00}$. In this manner,
$\bar{g}_{00}=1$ and $\bar{g}_{ij}=g_{ij}/g_{00}$ (Euclidean optical metric).

For the one-loop partition function it is possible to show that
\beq
\bar{Z}=J[g,\bar{g}]\,Z
\:,\eeq
where $J[g,\bar{g}]$ is the Jacobian of the conformal transformation.
Such a Jacobian can be explicitely computed \cite{dowk89-327-267},
but here we shall need only its structural form.
Using $\zeta$-function regularization for the determinant of the
differential operator we get
\beq
\ln Z=\ln\bar Z-\ln J[g,\bar g]
=\frac{1}{2}\ze'(0|\bar L_4\ell^2)-\ln J[g,\bar{g}]
\:,\label{lnZ-Zbar}\eeq
where $\ell$ is an arbitrary parameter necessary to adjust the
dimensions and $\ze'$ represents the derivative
with respect to $s$ of the function
$\ze(s|\bar L_4\ell^2)$ related to the operator $\bar L_4$,
which explicitly reads
\beq
\bar L_4=e^{-\si}L_4e^{-\si}=
-\partial_\tau^2-\bar\lap_3+\fr16\bar R
+e^{-2\si}\aq m^2+(\xi-\fr16)R\cq
=-\partial_\tau^2+\bar L_3
\:.\label{aconf}\eeq

The same analysis can be easily extended to the finite temperature
case \cite{dowk88-38-3327}.
Since the Euclidean metric is $\tau$ independent, one obtains
\beq
\ln\bar Z_\be&=&-\frac{\be}{2}\aq
\PP\ze(-\fr12|\bar L_3)
+(2-2\ln2\ell)\Res\ze(-\fr12|\bar{L}_3)\cq\nn\\
&&\hs+\lim_{s\to0}
\frac{d}{ds}\frac{\be}{\sqrt{4\pi}\Ga(s)}
\sum_{n=1}^\ii\int_0^\ii t^{s-3/2}\,e^{-n^2\be^2/4t}\,
\Tr e^{-t\bar{L}_3}\,dt
\label{lnZbeta}\:,
\eeq
where $\PP$ and $\Res$ stand for the principal part and the residue
of the function and one has to analytically continue before taking the
limit $s\to0$.
As usual, in the definition of $\zeta$-function
the subtraction of possible zero-modes of the corresponding operator
is left understood.
Of course, if the function $\ze(s|\bar L_3)$ is finite for $s=-1/2$,
the first term on the right-hand side of the latter equation is
just $-\frac\be2\ze(-\frac12|\bar{L}_3)$.
The latter formula is rigorously valid for a compact manifold.
In the paper we shall deal with
a non compact manifold, but nevertheless we shall make  a formal
use of this general formula, employing $\ze$-function associated with
continuum spectrum.

The free energy is related to the canonical partition function
by means of the equation
\beq
F_\be=E_v+\hat{F}_\be =-\frac{1}{\be}\ln Z_\be
=-\frac{1}{\be}\at\ln\bar Z_\be-\ln J[g,\bar g]\ct
\:,\label{FE}\eeq
where $E_v$ is the vacuum energy while $\hat{F}_\be$ represents the
temperature dependent part (statistical sum).
It should be noted that since we are considering a static space-time,
the quantity $\ln J[g,\bar g]$ depends linearly on $\be$ and,
according to Eq.~(\ref{FE}), it gives contributions only to the vacuum
energy term and not to the entropy, which  may be defined by
\beq
S_\be=\be^2 \partial_\be F_\be \:,
\label{entropy}
\eeq
and for the internal energy we assume the well known thermodynamical relation
\beq
U_\be=\frac{S_\be}{\be}+ F_\be
\:.\label{u}
\eeq

\s{Scalar fields in  Reissner-Nordstr\"om space-time}
\label{S:reiss}

Let us apply this formalism to the case of a massive
 scalar field in the 4-dimensional Reissner-Nordstr\"om background.
Our aim is to compute the entropy of this field using the latter formula.
 We recall that the metric we are interested in reads
\beq
ds^2=-\at1-\frac{r_H}r\ct^2\,(dx^0)^2+
\at1-\frac{r_H}r\ct^{-2}\,dr^2+r^2\,d\Omega_{2}
\:,\label{bh}
\eeq
where we are using polar coordinates, $r$ being the radial one and
$d\Omega_{2}$ the $2$-dimensional spherical unit metric.
The horizon radius is $r_H=MG=Q$,
$M$ being the mass of the black hole, $G$ the Newton constant and $Q$
its charge. The Hawking temperature is zero because it is proportional to
the difference between the horizon radii, which coincide in the
extremal case.
{}From now on, for the sake of convenience we put $r_H=1$; in this way
all quantities are dimensionless; the dimensions could be easily
restored at the end of calculations.

It may be convenient to redefine the Schwarzschild coordinates $(x^0,r)$
by means
\beq
x'^0=x^0\:,\hs
\rho=\frac{(r-1)}{1-2(r-1)\ln(r-1)-(r-1)^2}
\:,\label{ro}
\eeq
where $r$ is implicitly defined by Eq.~(\ref{ro}) and has the
expansion, valid near the horizon $\rho \sim 0$
\beq
r\sim1+\rho+O(\rho^2\ln\rho)
\:.\label{456}\eeq
The optical metric $\bar{g}_{\mu\nu}=g_{\mu\nu}/(-g_{00})$, which is
conformally related to the previous one and appears as an
ultrastatic metric, reads
\beq
d\bar s^2=-(dx'^0)^2
+\frac1{\rho^4}
d\rho^2+\frac{G(\rho)}{\rho^2}\,d\Omega_{2}
\:,\label{OM}\eeq
where
\beq
G(\rho)=\frac{r^4}{1-2(r-1)\ln(r-1)-(r-1)^2}
\sim1+2\rho\ln\rho+4\rho+O(\rho^2\ln\rho)
\label{67}\:.\eeq
Furthermore one has $\bar R=O(\rho^4\ln\rho)$.
In order to perform explicit computations, we shall consider
the large mass limit  of the black hole and the region near the
horizon. This leads to the approximated metric
\beq
d\bar s^2\sim-(dx'^0)^2
+\frac{1}{\rho^4} d\rho^2+ \frac{1}{\rho^2}d\Omega_{2}
\label{lsus1}
\:.\eeq
This can be considered as an approximation of the metric defined
by Eq.~(\ref{OM}) in the sense that, near the horizon $\rho=0$,
the geodesics are essentially the same. Eq.~(\ref{lsus1}) defines
a manifold with vanishing curvature.
Then, according to Eq.~(\ref{aconf}), the relevant operator becomes
\beq
\bar L_3=-\bar\lap_3+m^2\rho^2
=\rho^4\partial_\rho^2+\rho^2(\lap_2+m^2)
\label{barLN}
\:,\eeq
while the invariant measure reads
\beq
d\bar V=\rho^{-4}\,d\rho\,dV_{2}\:,
\eeq
where $\lap_2$ is the Laplace-Beltrami operator on the
unitary sphere $S^2$ and $dV_{2}$ its invariant measure.

The eigenfunctions of the operator $-\lap_2+m^2$ are the
spherical harmonics $Y_l^m(\te,\ph)$ and the eigenvalues
$\la_l^2=l(l+1)+m^2$.
Let $\Psi_{\la lm}=\phi_{\la l}(\rho)Y_l^m(\te,\ph)$
be the eingenfunctions of $\bar L_3$ with eigenvalues $\la^2$.
The differential equation which determines the related continuum spectrum
turns out to be
\beq
\aq\rho^4\,\partial_\rho^2-\rho^2 \la_l^2+\la^2
\cq\phi_{\la l}(\rho)=0
\:.\label{Bessel}
\eeq
The only solutions
with the correct decay properties at infinity
are the Bessel functions and we have
\beq
\phi_{\la l}(\rho)=\sqrt\rho
J_{\nu_l}(\rho^{-1}\la)
\:,\label{10}
\eeq
where $\nu_l^2=(l+\frac12)^2+m^2$ has been set.
For any suitable function $f(\bar L_3)$, we may write
\beq
<x|f(\bar L_3)|x>
=\int_0^\ii f(\la^2)\sum_{lm}\mu_l(\la)
Y_l^{*m}(\te,\ph)\phi^*_{\la l}(\rho)
Y_l^m(\te,\ph)\phi_{\la l}(\rho)\,d\la
\:,\label{hk}
\eeq
where $\mu_l(\la)$ is the spectral measure (density)
associated with the continuum spectrum.
It may be defined by means of equation
\beq
\at\phi_{\la l},\phi_{\la' l}\ct=
\frac{\de(\la-\la')}{\mu_l(\la)}
\:.\eeq
Using the asymptotic behaviour of the Bessel functions at
infinity \cite{grad80b} one can  show that
\beq
\mu(\la)\equiv\mu_l(\la)=\la
\:.\label{v2}
\eeq
This is the spectral measure associated with the Hankel inversion
formula \cite{chee83-18-575}. As a consequence, the heat kernel of
$\bar L_3$ may be written as
\beq
K_t(x|\bar L_3)=\int_0^\ii
d \la \la e^{-t \la^2}
\sum_{l=0}^\ii\frac{(2l+1)}{4\pi}\:
\rho\:J_{\nu_l}^2(\rho^{-1}\la)\,
\:.\eeq

To go on, we use a method based on the Mellin-Barnes representation.
In fact, we have \cite{grad80b}
\beq
\sum_{l=0}^\ii \frac{(2l+1)}{4\pi}
\:\rho\:J_{\nu_l}^2(\rho^{-1}\la)
=\frac{(4\pi)^{-3/2}}{\pi i}\int_{\Re s=3/2+c}
\frac{\Ga(s-\frac12)}{\Ga(s)}
\,\rho^{3-2s} \la^{2s-2}\,I(s)\,ds
\:,\label{MBR}
\eeq
where $c$ is a positive number smaller than $m^2$ or
than 1, according to whether one is considering the massive or the
massless cases respectively. We have introduced the function
\beq
I(s)=\sum_{l=0}^\ii(2l+1)
\frac{\Ga(\nu_l-s+1)}{\Ga(\nu_l+s)}\,,
\eeq
which is certainly convergent for $\Re s=3/2+c$, since for large $l$
one has \cite{olve74b}
\beq
\frac{\Ga(\nu_l-s+1)}{\Ga(\nu_l+s)}
\sim\nu_l^{1-2s}
\sum_{j=0}^\ii G_j(s)\nu_l^{-j}
\equiv \nu_l^{1-2s} Q(s,\nu_l)
\label{I}
\:,\eeq
$G_j(s)$ being computable polynomials in $s$.
For example we have
\beq
G_0(s)=1\:,\hs G_1(s)=0\:,\hs
G_2(s)=\frac{s(s-1)(s-1/2)}3
\:.\label{com}\eeq
Eq.~(\ref{I}) is useful in order to obtain the analytical continuation of
$I(s)$ for $\Re s<3/2$. In the massive case, the result is
\beq
I(s)=\sum_{j=0}^\ii G_j(s)f(s+\fr{j-1}2)
+\mbox{ analytical part}
\label{I1}
\:,\eeq
where
\beq
f(z)=\sum_{l=0}^\ii(2l+1)\nu_l^{-2z}
\:,\label{f}
\eeq
while, in the massless case
\beq
I(s)=\frac{\Ga(\frac32-s)}{\Ga(\frac12+s)}+
2\sum_{j=0}^\ii G_j(s)\ze_H(2s+j-2;\fr12)
+\mbox{ analytical part}
\label{I2}
\:,\eeq
$\ze_H(z;q)$ representing the Riemann-Hurwitz zeta-function, which has
only a simple pole at $z=1$, with residue equal to 1.
In the same way, the function $f(z)$ admits an analytical continuation
in the whole complex $z$ plane with a simple pole at $z=1$
and residue equal to 1. This means that in both the cases,
for $\Re s<3/2+c$, $I(s)$ is a meromorphic function with possible
simple poles at $s=\frac{3-j}2$ ($j=0,1,...$). However we observe that
only $s=3/2$ is a true pole, with residue equal to 1, since
the residues $G_j(\frac{3-j}2)$ ($j\geq1$) in all other possible
poles are equal to 0.
This can be easily seen by observing that for integer or
half-integer $s=\pm\frac{N}2$ ($N\geq0)$ we have
\beq
Q(-\fr{N}2,\nu_l)=\prod_{0\leq k=odd/even}^N
	\at1-\frac{k^2}{4\nu^2}\ct\:,
\hs\mbox{for odd/even }N\,\label{ff1}
\eeq
\beq
Q(\fr{N}2,\nu_l)=\prod_{0\leq k=odd/even}^{N-2}
\at1-\frac{k^2}{4\nu^2}\ct^{-1}\:,
\hs\mbox{for odd/even }N
\:.\label{ff2}\eeq
{}From equations above, we see that $Q(\pm\frac{N}2,\nu_l)$
is an even function in $\nu_l^{-1}$ and this means that the
corresponding $G_j$ are vanishing if $j$ is odd. Moreover,
from Eq.~(\ref{ff1}) we see that $Q(-\frac{N}2,\nu_l)$ is a
polynomial of degree $N+1$ or $N$ according to whether
$N$ is odd or even. Then $G_j({\frac{3-j}2})\equiv0$
if $3-j\leq0$. The coefficients for $s=0$, $\frac12$ and 1
can trivially computed using Eq.~(\ref{ff2}). They all are
vanishing but $G_0$, which is equal to 1 for any $s$.
As a result, $I(s)$ is analytic for $\Re s<3/2$.

The heat kernel reads
\beq
K_t(x|\bar L_3)=\frac{(4\pi)^{-3/2}}{2\pi i}
\int_{\Re s=3/2+c}
\Ga(s-\fr{1}{2})
\,\rho^{3-2s}\,t^{-s} \,I(s)\,ds
\:.\label{bn}
\eeq
Since we are mainly interested in global quantities, like the
partition function, we integrate Eq.~(\ref{bn}) over
$\bar\M^3$, paying attention to the fact that the integration
over $\rho$ formally gives rise to horizon divergences.
In order to regularize such divergences,
we introduce a horizon cutoff $\ep>0$.
At the end of the computation, we shall take the limit $\ep\to0$.
In this way we have
\beq
\Tr e^{-t\bar L_3}(\ep)
=\frac{A(4\pi)^{-3/2}}{4\pi i}
\int_{\Re s=3/2+c}\frac{\Ga(s-\fr{1}{2})}{s}\,
\ep^{-2s}\,t^{-s} \,I(s)\,ds
\label{4bh1}
\:,\eeq
where the horizon area $A$ is equal to $4\pi r_H^2$ and the
integration in $\rho$ has been performed.

Now we can shift the vertical contour to the left hand
side in the complex plane. There are simple poles at all the
half-integers $s\leq3/2$ and also in $s=0$.
Since we have to take the limit $\ep\to0$,
only the non negative poles will give non vanishing (divergent)
contributions to the integral. As a result
\beq
\Tr e^{-t\bar L_3}(\ep)=
\frac{A}{(4\pi t)^{3/2}}\aq
\frac1{3\ep^3}+\frac{I(\frac12)}{\ep}\,t
-\sqrt\pi I(0)\,t^{3/2}\cq
\:.\eeq
It should be noted the appearance of a horizon divergence
more severe than the one which one has in the non extremal case.

The corresponding partition function may be evaluated making use
of Eq.~(\ref{lnZbeta}).
We have
\beq
\ln Z_\be=-\be F_\be &=&
 \be A\aq j_\ep-\frac12\ze(-\fr12|\bar L_3)\cq
\nn\\&&\hs
+\frac{\be A}{(4\pi)^2}\aq
\frac{8\pi^4}{135\ep^3\be^4}
+\frac{2\pi^2I(\frac12)}{3\ep\be^2}
+\frac{2\pi I(0)}{\be}\ln\be\cq
\:,\label{bhf}
\eeq
where we have written the Jacobian contribution
due to the conformal transformation in the form $A\be j_\ep$.
The horizon divergences which appear in Eq.~(\ref{bhf}) are also contained
in the Jacobian factor, while ultraviolet divergences are present in
the vacuum energy part.
As is well known, one needs a renormalization
in order to remove the vacuum divergences. We recall that these
divergenges, as well as the Jacobian conformal factor, being linear in $\be$,
do not make contributions to the entropy.
As a consequence, the first quantum corrections at temperature $T=1/\be$
to entropy read
\beq
S_\be=A\aq\frac{2\pi^2}{135\ep^3\be^3}
+\frac{I(\frac12)}{12\ep\be}
+\frac{I(0)}{8\pi}\at\ln\be-1\ct\cq
\:.\eeq

The functions $I(1/2)$ and $I(0)$ can be evaluated in the sense of
analytical continuation. In the massive case we have
\beq
I(\fr12)=f(0)=2\ze_H(-1|\fr12)=\frac1{12}
\:,\label{mass}
\eeq
\beq
I(0)=f(-\fr12)=-\frac1{\sqrt\pi}\sum_{n=0}^\ii
(-1)^n m^{2n}\Ga(n-\fr12)\,\ze_H(2(n-1)|\fr12)
\:,\label{mass1}
\eeq
while for the massless one, $I(\frac12)$ is again given by
Eq.~(\ref{mass}), but $I(0)$ vanishes.

Then we have the final result (restating $r_H$ by dimensional
arguments)
\beq
S_\be=A\aq\frac{2\pi^2 r_H^4}{135\ep^3\be^3}
+\frac1{144\ep\be}
+\frac{f(-\frac12)}{8\pi}\at\ln\fr{\be}{r_H}-1\ct\cq
\:,\eeq
\beq
S_\be=A\aq\frac{2\pi^2 r_H^4}{135\ep^3\be^3}
+\frac1{144\ep\be}\cq
\:,\eeq
valid for massive and massless fields respectively.
The leading term of above equations is in agreement
with Refs.~\cite{mitr95u-42,deal95u-33}.

\s{Concluding remarks}
\label{S:conc}

In this paper, making use of conformal transformation techniques, we
have investigated in some detail, the first quantum corrections to an
extremal Reisnerr-Nordstr\"{o}m black hole. Our approach does not make use
of brick-wall boundary conditions, but the horizon cut-off  is
introduced in the computation of the heat-kernel trace. Although the
spectrum of the relevant operator is continuous, the
computation of the statistical sum can be done, in the large mass
limit of the black hole and the leading
divergences of the entropy are in agreement with the ones obtained in
Refs. \cite{mitr95u-42} (brick-wall method) and
\cite{deal95u-33}, in which the optical manifold method is implemented
in an alternative way.
However, we would
like to stress a technical point. In the case of canonical horizons,
the optical spatial section turns out to be smooth, for example in the
Rindler space-time, it concides with the hyperbolic space. In the
case of non canonical horizons, the optical spatial section has
conical singularities and this fact leads to the appearance of a
linear divergence besides the leading cubic one.

As we have already recalled, in this extremal case the Hawking temperature
vanishes, but
it has been recently observed that the corrresponding Euclidean
solution may be identified with an arbitrary $\be$
\cite{hawk94u-13,teit94u-3}. As a result, in
the extreme case, one has a quantum divergent entropy which does not
satisfy the area law. Since our techniques can be applied to every
space-time with non canonical horizons, we may conclude that this area law
violation holds also in these cases. We note, however, that the area
law violation is also present in the extremal dilatonic black hole
\cite{mitr95u-42}, where the linear divergence is vanishing in the
extremal case.

Finally few words about these horizons divergences. In the case of
canonical horizons, it has been recently proposed to cure the linear and
logarithmic divergences by standard one-loop renormalization of the Newton
constant \cite{suss94-50-2700,solo95-51-609,furs94u-20,solo95u-22} within the
formalism of the quantum gravity effective action (for a recent review
on this issue see, for example,
\cite{buch92b}). However, apart the criticism of Ref.
\cite{barb95u-155}, it is not
clear to us whether the same procedure may be applied to the case of
space-times with non canonical horizons, since here, as we have shown,
the leading horizon divergence is cubic.

On the other hand, in the non
extremal case, one may choose the horizon cutoff of the order of the
Planck length ($\ep^2=G$), in order to reproduce the area law
\cite{thoo85-256-727,frol94u-36}. In the extremal case, this seems
problematic, since there is no natural value for $\be$. Alternatively,
for this problem, one may try to make use of string theory, which may
be considered a perturbatively ultaviolet finite model for quantum
gravity (see for example Refs.
\cite{suss94-50-2700,dabh94u-68,lowe94u-215,empa94u-3,eliz94u-24}).

\end{document}